# Development of radioactive beams at ALTO: Part 2. Influence of the UC$_x$ target microstructure on the release properties of fission products

Julien Guillot[a*], Brigitte Roussière[a], Sandrine Tusseau-Nenez[b], Denis Grebenkov[b], Nicole Barré-Boscher[a], Elie Borg[a], Julien Martin[a]

[a] Institut de Physique Nucléaire d'Orsay CNRS/IN2P3 UMR 8608 - Université Paris Sud - Université Paris Saclay, F-91406 ORSAY Cedex
[b] Laboratoire de Physique de la Matière Condensée, CNRS, Ecole Polytechnique - Université Paris Saclay, Route de Saclay, F-91128 Palaiseau

*Corresponding author: guillotjulien@ipno.in2p3.fr, tel: +33169155101

___________________________________________________________________________

___________________________________________________________________________

**Abstract**:

Producing intense radioactive beams, in particular those consisting of short-lived isotopes requires the control of the release efficiency. The released fractions of 11 elements were measured on 14 samples that were characterized by various physicochemical analyses in a correlated paper (Part 1). A multivariate statistical approach, using the principal component analysis, was performed to highlight the impact of the microstructure on the release properties. Samples that best release fission products consist of grains and aggregates with small size and display a high porosity distributed on small diameter pores. They were obtained applying a mixing of ground uranium dioxide and carbon nanotubes powders leading to homogeneous uranium carbide samples with a porous nanostructure. A modelling under on-line ALTO conditions was carried out using the FLUKA code to compare the yields released by an optimized and a conventional target.

___________________________________________________________________________

Introduction:

The research and development program at the ALTO facility (Accélérateur Linéaire et Tandem d'Orsay) of the Institut de Physique Nucléaire d'Orsay (IPNO) aims to provide new beams of exotic and neuton-rich nuclei, as intense as possible. At ALTO, neutron-rich nuclei are produced by photofission in thick uranium carbide targets. An efficient way to improve the intensities of the radioactive beams, in particular those consisting of short-lived isotopes, would be to develop dense and porous targets. These two properties are antagonistic but essential to increase the fission fragment quantity produced and diffuse them out of the target respectively.

Much research work has been done to improve release efficiencies by studying the influence of several parameters. Numerous studies have pointed out the impact of the uranium precursors [1]–[5] or lanthanum precursors taken as uranium structural equivalent [6], [7] on the target porosity. It appears that, as the precursors generate more gas, the porosity is favoured. Whatever the metal precursor used for the synthesis of uranium carbide (UC$_x$) [4], [8]–[11], [14]–[16], [22], lanthanum carbide (LaC$_x$) [12], [13], [17], [18], [21] or titanium carbide (TiC) [19], [20] targets, the carbon source (graphite [8]–[13], reticulated vitreous carbon foam [14]–[16], graphene [17], [18], carbon black [19], [20] and carbon nanotubes [4], [19]–[22]) is also a key point because this parameter impacts not only the amount of porosity but also the carbide grain size and the sintering step. Finally, the ratio of metal precursor (uranium) to carbon must be a fine balance between obtaining high release efficiency and sufficient production of fission products [23]–[27]. Moreover some recent studies suggest controlling the porosity by adding polymer sacrificial particles which will leave a footprint after degassing due to the heating [28]–[30]. Studies on calcium oxide (CaO) ISOL targets also showed the importance of the temperature on the target microstructure in order to control the sintering and the grain size [31], [32]. It was recently suggested that the reaction rate velocity during carburization could also be a point to be optimized [33]. In addition to experimental approaches [34]–[36], the release dependence upon grain size and porosity has been studied for various isotopes in $UC_x$ targets by simulations carried out with the RIBO code [37], [38]. Finally, recent results on $UC_x$ targets have shown that using refractory targets with a nanometric structure enables the production of new and more exotic radioactive isotope beams [22]. The correlated paper of this study (Part 1) gives more details on the contribution of the studies cited here on the target synthesis [39].

In this context, we have undertaken to determine the impact of the physicochemical characteristics of a target on its release properties. As a first step, seven samples ranging from a dense UC sample to a highly porous $UC_x$ one elaborated with carbon nanotubes were studied. Open porosity appears to play the most significant role and the use of carbon nanotubes proves very promising provided the microstructure remains stable when the sample is heated at high temperature for a long time [4]. In a second step, after having investigated the parameters involved in the target manufacturing (grinding and mixing of precursor powders, pressing, carburization and dwelled time at high temperature), we produced fourteen samples from different uranium and carbon sources applying three mixing protocols. An exhaustive physicochemical description of the samples was obtained, and strong correlations were established between the methods of sample preparation and the structural properties, such as the quantity of porosity, the pore-size distribution, the size of grains and aggregates [39]. In the present paper, the release properties of these fourteen samples are investigated. The samples were irradiated with a deuteron beam delivered by the tandem accelerator of the laboratory and the released fraction of 11 elements was measured by gamma spectrometry. A statistical analysis of the results using the Principal Component Analysis (PCA) method allowed us to establish strong correlations between the release properties of the samples and some of their structural properties.

1. Experimental protocol
1.1. Description of the samples

Two uranium (uranium dioxide and uranium oxalate) and three carbon (graphite, carbon nanotube-CNT), graphene) sources were used to produce 14 batches of samples according to 3 different mixing protocols labelled conventional protocol (CP), developing protocol (DP) and graphene protocol (GP). 0.4 g and 1 g were taken from the powder mixtures containing the CNT and graphite (or graphene) respectively. Each powder sample was pressed 6 seconds at 220 MPa to form a green pellet with about 13 mm diameter and 1.7 mm thickness. All the pellets were carburized at 1800 °C for 2 hours and pellets No. 9, 10, 11 and 12 underwent an additional heat treatment at 1800 °C for 12 days. Despite the care taken during the preparation, after heating, the samples show significant differences in diameter and thickness (Table 1).

Table 1: Mass ($m$), diameter ($d$) and thickness ($t$) of the pellets used for the release experiment. The uranium quantity contained in each pellet is given in column 8 and the intensity of the deuteron beam used during the irradiation in column 9. Samples are labelled by the uranium and carbon sources and the mixing protocol used (for details, see ref. [39]). The PARRNe samples correspond to pellets used as conventional target at ALTO and build of $UO_2$ ground and graphite according to the conventional protocol.

| No./Samples | P1 | | | P2 | | | Uranium quantity per pellet ($\times 10^{-3}$ mol) | Beam intensity I (nA) |
|---|---|---|---|---|---|---|---|---|
| | m (g) | d (mm) | t (mm) | m (g) | d (mm) | t (mm) | | |
| No.1 $UO_2$ ground + CNT CP | 0.32 | 13.52 | 2.24 | 0.33 | 13.50 | 1.58 | 1.0 | 50 |
| No.2 $UO_2$ ground + CNT DP | 0.33 | 12.37 | 1.62 | 0.33 | 12.38 | 1.61 | 1.0 | 50 |
| No.3 $UO_2$ ground + graphene GP | 0.75 | 11.62 | 1.80 | 0.77 | 11.56 | 2.00 | 2.4 | 20 |
| No.4 OXA + graphite CP | 0.32 | 9.61 | 1.29 | 0.32 | 9.62 | 1.33 | 1.0 | 50 |
| No.5 OXA ground + CNT DP | 0.21 | 11.28 | 1.58 | 0.21 | 11.31 | 1.57 | 0.7 | 50 |
| No.6 OXA + CNT DP | 0.22 | 12.39 | 1.63 | 0.22 | 12.40 | 1.63 | 0.7 | 50 |
| No.7 PARRNe BP894 | 0.82 | 12.62 | 1.43 | 0.84 | 12.70 | 1.45 | 2.6 | 20 |
| No.8 PARRNe BP897 CP | 0.82 | 12.35 | 1.81 | 0.82 | 12.34 | 1.80 | 2.6 | 20 |
| No.9 PARRNe BP897 CP 12d | 0.80 | 12.31 | 1.81 | 0.81 | 12.29 | 1.79 | 2.6 | 20 |
| No.10 $UO_2$ ground + CNT CP 12d | 0.33 | 12.69 | 2.05 | 0.33 | 12.68 | 1.90 | 1.0 | 50 |
| No.11 UO2 ground + CNT DP 12d | 0.33 | 12.54 | 1.66 | 0.33 | 12.55 | 1.62 | 1.0 | 50 |
| No.12 UO2 ground + graphene GP 12d | 0.77 | 11.52 | 1.79 | 0.78 | 11.46 | 1.81 | 2.5 | 20 |
| No.13 UO2 ground + CNT-5moles DP | 0.33 | 11.01 | 1.26 | 0.33 | 11.02 | 1.24 | 1.1 | 50 |
| No.14 UO2 ground + CNT-7moles DP | 0.33 | 13.02 | 1.95 | 0.33 | 13.08 | 1.98 | 1.0 | 50 |

The protocols used to synthesize the samples and the techniques and methods implemented to characterize them are detailed in Guillot *et al.* [39] and reported in complementary data. Each sample was thus described by 14 variables: the proportion of the $UC_2$, $UC$ and $C$ phases, the size of the $UC$ and $UC_2$ crystallites, the $UC_x$ grain size, the $UC_x$ aggregate size, the percentages of open and close porosities and the distribution of the open porosity on pores with 0.035, 0.2, 3, 10 and 30 µm average diameters (see Table 2 of ref. [39] and complementary data).

1.2. Release measurement

The released fractions were determined from the amount of fission products remaining in the samples after heating. Fission products were obtained from the uranium (99.75 wt% of $^{238}U$ and 0.25 wt% of $^{235}U$) fission induced by the neutron flux generated by stopping a 26 MeV deuteron beam in a graphite converter. The measurements were performed following the procedure developed by Hy *et al.* and Tusseau-Nenez *et al.* [2], [4]. Improvements have been made to the pellet positioning in the specimen holder and to the heating device in order to increase the reliability and the reproducibility of the measurements. They are described below.

Among all the pellets synthesized for each sample type, two, called $P_1$ and $P_2$ whose the dimensions are given in Table 1, were selected to perform the release measurements. The first step consisted in irradiating the two pellets placed in the specimen holder, one behind the other and both behind a graphite converter. They were immobilized by graphite centering ring and shim in order to be perfectly in the beam axis. The deuteron-beam intensity impinging on the converter was determined from the uranium quantity contained in the samples. As shown in Table 1, pellets containing carbon nanotubes contain less uranium than pellets containing graphite or graphene. They were irradiated with a higher-intensity beam in order to obtain, for all samples, very close radioactivity levels and thus to carry out the γ-spectroscopy measurements with very similar temporal (irradiation, waiting and counting times) and spatial (source-detector distance) conditions. All samples were irradiated for 20 minutes. For radiation protection and safety reasons, a waiting time of the order of 30 minutes was required before removing the pellets from the sample holder.

The second step consists in placing each pellet in front of a Ge(HP) detector in order to perform a first γ-measurement aiming to compare the amount of fission products created in the two pellets.

In the third step, the pellet $P_1$ was heated to 1768 °C for 30 minutes. The temperature of 1768 °C corresponds to the temperature used in our previous R&D experiments at ALTO. Using this temperature, we can check the reproducibility of the experiments and validate the protocol used for the study of the fission-product release. The pellet was placed between two tantalum crucibles as shown in Figure 1. Compared to what was done in ref. [4], the confinement is increased and the temperature is kept more homogeneous. Then the oven chamber was closed and pumping started. The heating step was no more manually operated as before but was regulated by a programmable logic controller, which ensures the reproducibility of the procedure. As soon as the pressure reached $10^{-2}$ Pa, the controller started the heating cycle: a ramp of 34 A/min was applied until 340 A was reached. This current value corresponds to a temperature of 1768 °C checked by the fusion of a small platinum wire and maintained for 30 minutes, time set for the diffusion of the fission products out of the pellet. The pressure in the chamber was $8.5 \times 10^{-4}$ Pa during heating. After this heating time, the controller lowered the current by applying a 50 A/min ramp. The pumping was then stopped and when the pressure in the chamber rose to $10^{-2}$ Pa, argon, and not air because of the uranium carbide pyrophoricity [40], was introduced. Due to the oven inertia, returning to a temperature below 70 °C takes 14 minutes.

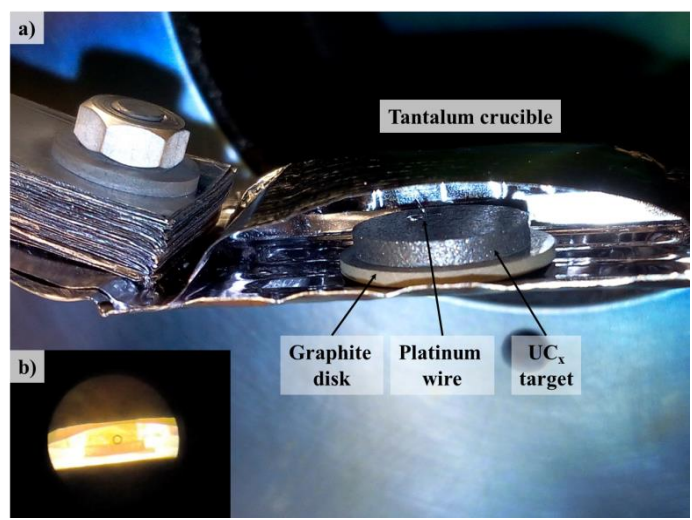

Figure 1: a) oven consisting of two tantalum crucibles, in order to maximally confine the $UC_x$ pellet placed on a graphite disk b) oven during the heating experiment.

The last step was to perform a one-hour γ-measurement of the pellets which aims to determine the amount of fission products remaining after heating.

The waiting time between the end of the irradiation and the start of the first γ-spectroscopy measurement was 47 ± 8 min. The time of this first measurement was 10 min and the delay between the end of this first counting and the beginning of the second one was 62 ± 3 min. The very low dispersion (value that can be compared to the 18 min obtained in the previous experiment [4]) was achieved by automating the heating procedure.

2. Experimental results

All the γ-spectra were analysed with the Fityk software [41] and the released fractions (*RF*) could be determined for 11 elements: krypton (Kr), strontium (Sr), ruthenium (Ru), tin (Sn), antimony (Sb), tellurium (Te), iodine (I), cesium (Cs), barium (Ba), lanthanum (La) and cerium (Ce). The release properties of yttrium (Y) and xenon (Xe) were not evaluated, contrary to what was done in refs. [2], [4]. In this case, both elements are identified in the γ-spectra by only one γ-ray originating from the decay of $^{91m}Y$ and $^{135g}Xe$ respectively. But these latter nuclei are much less produced by fission than their radioactive precursors (200 times and 20 times less, respectively). In this case the released fraction measured is nothing more than an apparent value induced by the precursor release. Figure 2 shows the apparent released fraction calculated for $^{91m}Y$, assuming no direct Y production by fission and only Sr release, versus the released fraction measured. The red dots represent the results obtained for the 14 samples, they lie along the $RF_{apparent} = RF_{measured}$ black straight line, which confirms that the $^{91m}Y$ activity decrease observed is due to the release of $^{91}Sr$.

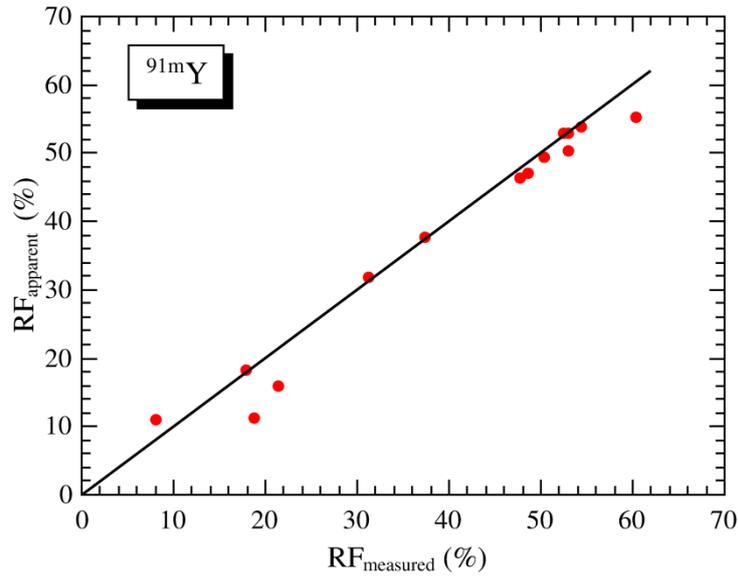

Figure 2: Apparent released fraction calculated for $^{91m}$Y versus the released fraction measured.

Dombsky *et al.* had shown the influence of the target thickness on the production of short-lived nuclides [42]. The released fractions extracted from the γ-spectrometry measurements have been corrected from the pellet thickness applying the method described in ref. [43]: all the *RF* values have been standardized on pellets with 13 mm in diameter and 1.7 mm in thickness. Figure 3 shows these results obtained for two "PARRNe[1] 894BP" samples irradiated by Tusseau-Nenez *et al.* [4] and in this study.

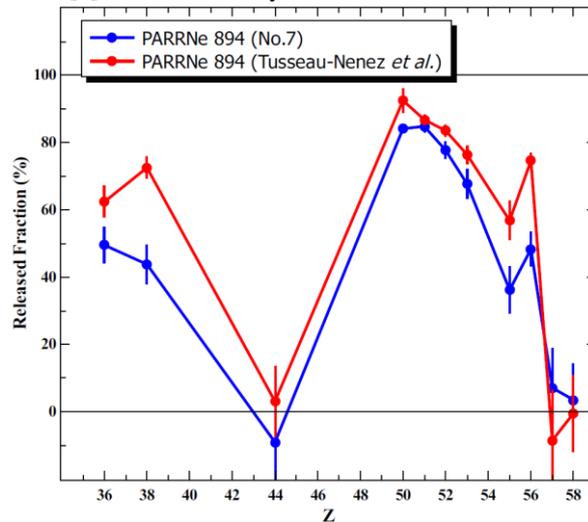

Figure 3: Standardized released fractions obtained for two "PARRNE 894BP" samples in 2016 and in this study.

The *RF* changes as a function of Z are very similar for the two "PARRNe 894BP" samples. However, the released fractions obtained during the experiment carried out by Tusseau-Nenez *et al.* [4] are systematically higher. This difference can originate from a slightly higher temperature than expected in the experiment described in ref. [4]. The temperature accuracy and reliability were difficult to control in the previous experiment. To overcome this drawback the heating procedure was controlled by a programmable device. To validate this hypothesis, the sample No. 8 was synthesized applying the same protocol as the one used for sample No. 7. Figure 4a shows that the samples No. 7 and No. 8 exhibit almost the same released fractions whatever the element, highlighting the high reproducibility of the temperature obtained with the programmable device.

---

[1] The PARRNe samples correspond to pellets used as conventional target at ALTO

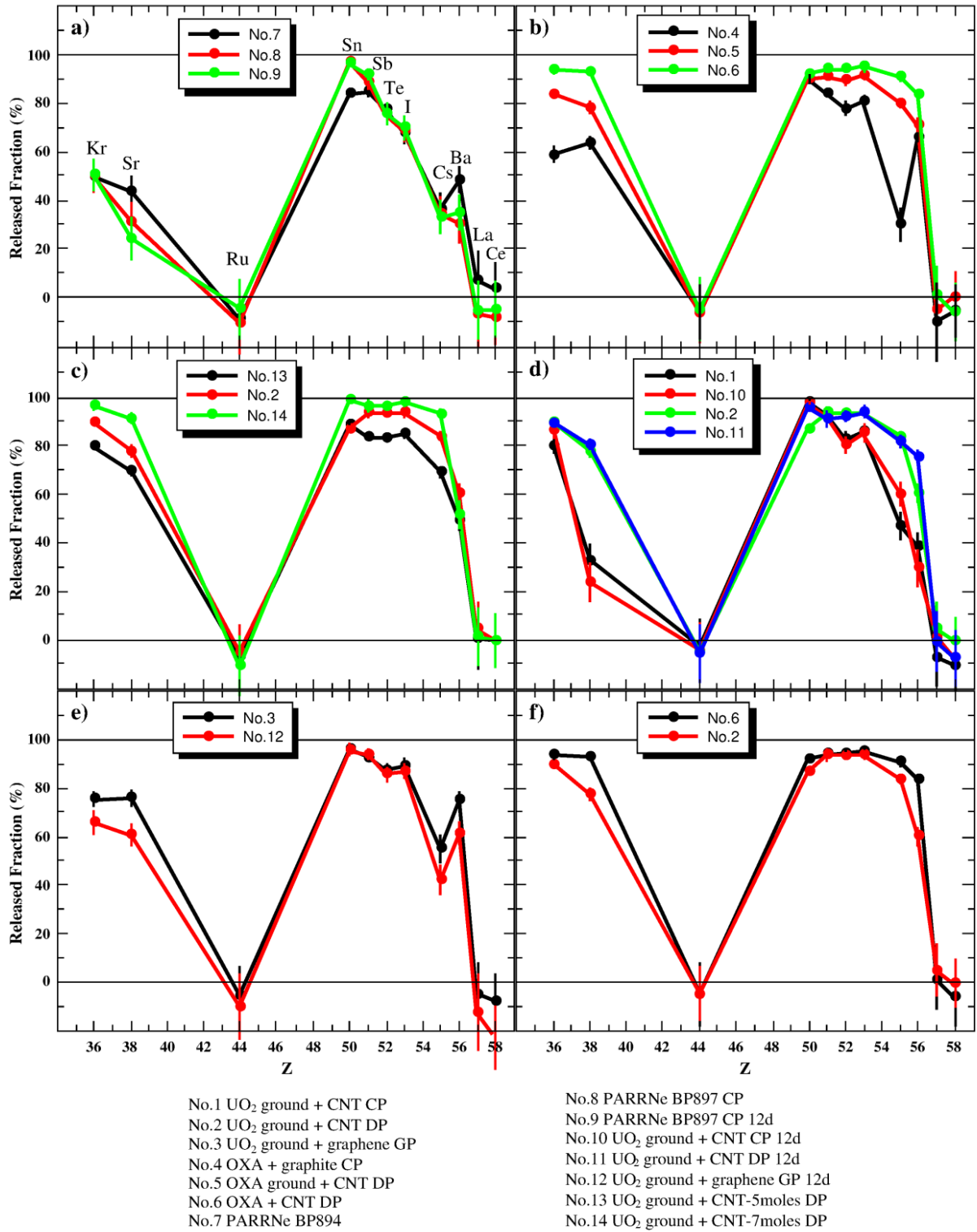

No.1 UO$_2$ ground + CNT CP  
No.2 UO$_2$ ground + CNT DP  
No.3 UO$_2$ ground + graphene GP  
No.4 OXA + graphite CP  
No.5 OXA ground + CNT DP  
No.6 OXA + CNT DP  
No.7 PARRNe BP894  

No.8 PARRNe BP897 CP  
No.9 PARRNe BP897 CP 12d  
No.10 UO$_2$ ground + CNT CP 12d  
No.11 UO$_2$ ground + CNT DP 12d  
No.12 UO$_2$ ground + graphene GP 12d  
No.13 UO$_2$ ground + CNT-5moles DP  
No.14 UO$_2$ ground + CNT-7moles DP  

Figure 4: Standardized released fractions obtained for the 14 samples.

Figure 4a also shows the *RF* values obtained for the PARRNe sample No. 9, the third one prepared with the conventional protocol (CP). The superposition of the released fractions for the No.8 and No.9 samples demonstrates that the heat treatment at 1800 °C for 12 days induced no change in the sample release properties. Figure 4b illustrates the influence of the carbon source on the release properties of samples made from uranium oxalate. The use of carbon nanotubes (samples No. 5 and No. 6) improves the release for six elements: Kr, Sr, Sb, Te, I and Cs. A slight decrease in the released fraction is observed, for pellets containing CNT, when the uranium oxalate was ground (sample No. 5). Contrary to $UO_2$, the grinding of uranium oxalate does not improve the fission-product release. The grinding step, having modified the uranium oxalate structure, could generate a different granular microstructure limiting the diffusion of the fission products.

Figure 4c compares samples with different C/U molar ratios: 5, 6 and 7 for samples No. 13, 2 and 14 respectively. It appears clearly that excess of carbon improves the release properties. Figure 5 shows the standardized *RF* for two samples synthesized from uranium oxalate and graphite but with different molar ratios C/U (6 and 3) and confirms the benefit of carbon excess.

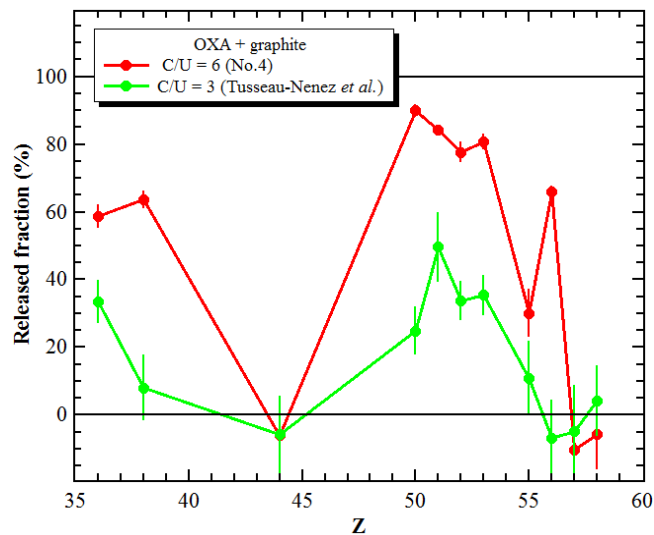

Figure 5: *RF*s for two samples synthesized from uranium oxalate and graphite but with different molar ratios C/U (6 and 3).

Figure 4d shows the influence of the precursor-powder mixing method on the release properties. The 4 samples were prepared by mixing ground $UO_2$ powder with carbon nanotubes according to the conventional protocol (CP, samples No. 1 and 10) or to the developing protocol (DP, samples No. 2 and 11). Much better released fractions are obtained for samples No. 2 and 11, that were elaborated following the method leading to very homogeneous mixtures [39]. As in the case of PARRNe samples and whatever the method of mixing, the released fractions are not reduced by long-term heating.

Figure 4e shows the released fractions obtained for the samples No. 3 and 12 prepared according to the graphene protocol (GP). In this case, the long-term heating results in a slight decrease of the release properties for Kr, Sr Cs and Ba that can be explained by the weak grain growth occurring in both samples [39]. However, it is worth noting that the released fractions observed for the samples prepared with the GP protocol are higher than those obtained for the PARRNe samples elaborated with graphite (see Fig. 4a and 4e).

Figure 4f illustrates the influence of the uranium source on the release properties of samples made from CNT with the DP protocol. The released fractions measured are better with uranium oxalate (sample No. 6) than with $UO_2$ (sample No. 2) for two elements, Sr and Ba. This result can be explained by a higher open porosity in the sample No. 6 (74 % instead of 68 %).

Based on these results and previous ones, the open porosity clearly appears to be a determining factor for high releases. This porosity is correlated to the use of CNT instead of graphite or to the chosen C/U ratio. Through these experimental studies, the role of the pore size distribution and the $UC_x$ grain size in the release efficiency have to be clarified.

3. Statistical analysis by Principal Component Analysis

As in the case of the physicochemical analysis of samples [39], a principal component analysis (PCA) was performed in order to highlight the correlations between the variables related to the release and those associated to the physicochemical properties. To carry out this PCA, each of the 14 samples was described by the 11 physicochemical variables statistically significant selected as active in ref. [39] as well as the $RF$ values measured for 6 elements. The $RF$ values for Ru, Sn, Sb, La and Ce were not considered as active variables because they do not exhibit any variability among samples. The temperature used, 1768 °C, is too low to release La, Ce and Ru. As for Sn and Sb, they are very well released by all the samples, therefore these five elements cannot be used to differentiate the samples. The result of this analysis is presented in Figure 6.

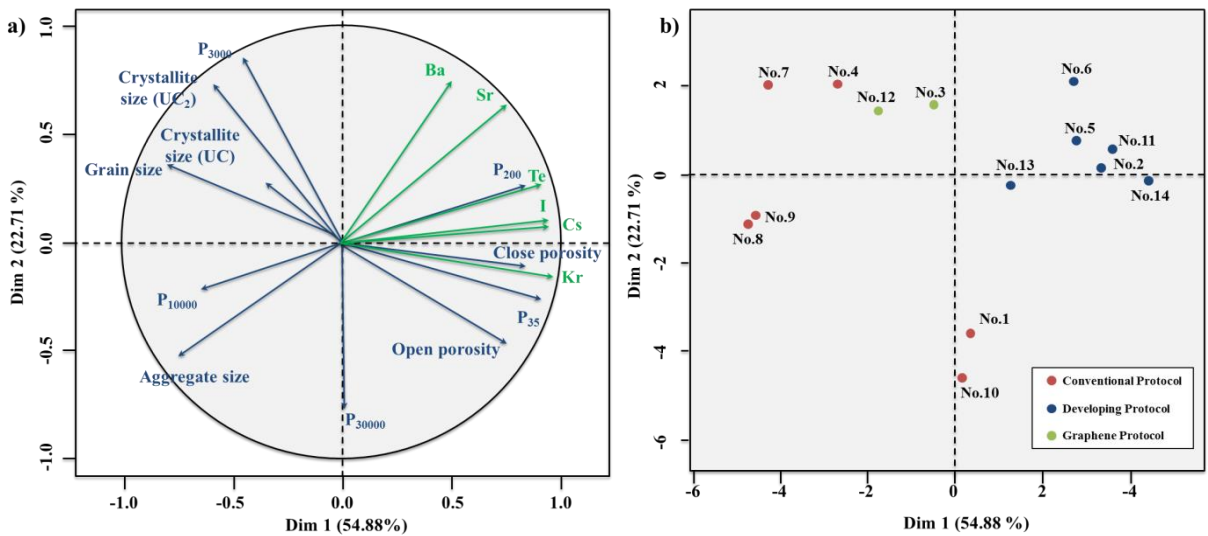

Figure 6: PCA analysis result a) the graph of the variables b) the graph of the samples according to the different protocols.

With 17 independent variables describing 14 individuals, the percentage of inertia in the first principal plane is expected to be 42.9 % [44]. The plane formed by the two principal components explains 77.6 % of the information contained in our data set. This high percentage of inertia shows that high correlations exist between the variables.

Figure 6a) shows that almost all variables are well described in the plane (1; 2) since they have a cos² (the quantity measuring the quality of the representation of a variable or of a sample) greater than 0.6, except the variables "crystallite size UC" and "$P_{10000}$" with cos² values equal to 0.2 and 0.4 respectively. The first principal component (Dim 1) is highly and positively correlated with the released fractions of Ba, Sr, Te, I, Cs and Kr, with the open and close porosities and with the pores with small diameters, 35 nm "$P_{35}$" and 200 nm "$P_{200}$". It is negatively correlated to the variables "aggregate size", "grain size" and "$UC_2$ crystallite size". The second principal component is positively correlated to the pores with a diameter of 3 μm ("$P_{3000}$"), the Ba and Sr released fractions and negatively to the pores with a diameter of 30 μm ("$P_{30000}$").

Figure 6b) depicts the 14 samples in the first principal plane. Almost all samples are well represented in this plane since their cos² values are higher than 0.7, except 3 of them, the samples No. 3, 12 and 13 which have cos² values equal to 0.4, 0.5, and 0.4, respectively.

The samples defined by a positive coordinate on Dim 1 lie on the right side of Fig. 6b). They are all made with carbon nanotubes. They are characterized by high $RF$ values for Kr, Te, I and Cs, by a high open porosity distributed on small pores ("$P_{35}$" and "$P_{200}$") and by small grain and aggregate size. Contrary to the samples No. 2, 5, 6, 11, 13 and 14, the samples No. 1 and 10 exhibit small $RF$ values for Sr and Ba and a high proportion of large pores ("$P_{30000}$"), two features associated with a negative value on Dim 2. Thus this latter axis allows sorting the samples according to the mixing protocol used: the conventional protocol for the samples No. 1 and 10 and the developing protocol for the samples No. 2, 5, 6, 11, 13 and 14. It is worth noting, that, although the sample No. 13 is not well represented in the plane (1, 2), it lies close to the samples synthesized with the same mixing protocol.

Samples with large grain and aggregate size, low open porosity and lower release properties appear in the left side of Fig. 6b) since these characteristics imply a negative coordinate on Dim 1. These are the samples No. 3, 4, 7, 8, 9 and 12. The samples No. 3 and 12, having physicochemical properties close to the PARRNe sample group but better *RF* values, have a greater coordinate on Dim 1. The sample No. 7 differs from No. 8 and 9 by an important proportion of pores with 3 μm diameter and higher *RF* values for Sr and Ba. This results in a positive coordinate on Dim 2 for the sample No. 7 and a negative one for No. 8 and 9.

In the PCA framework, qualitative variables, considered as supplementary variables, can be connected to the principal components by representing, in the factorial planes, the gravity centres of the individuals having the same modality. Figure 7 shows the representation obtained in the first principal plane for the qualitative variables "carbon source", "mixing protocol", "uranium source", "12 days heating", "C/U (for DP and CNT samples)" and "C/U (all samples)". The modalities of the variables "mixing protocol", "uranium source" and "carbon source" are well separated in the plane (1, 2), they differ therefore significantly on one or both dimensions defining this plane. On the contrary, the overlap between the two modalities of the "12-day heating" variable is maximum, which means that these two modalities are not significantly different; the physicochemical properties as well as the release ones are not changed by long-term heating.

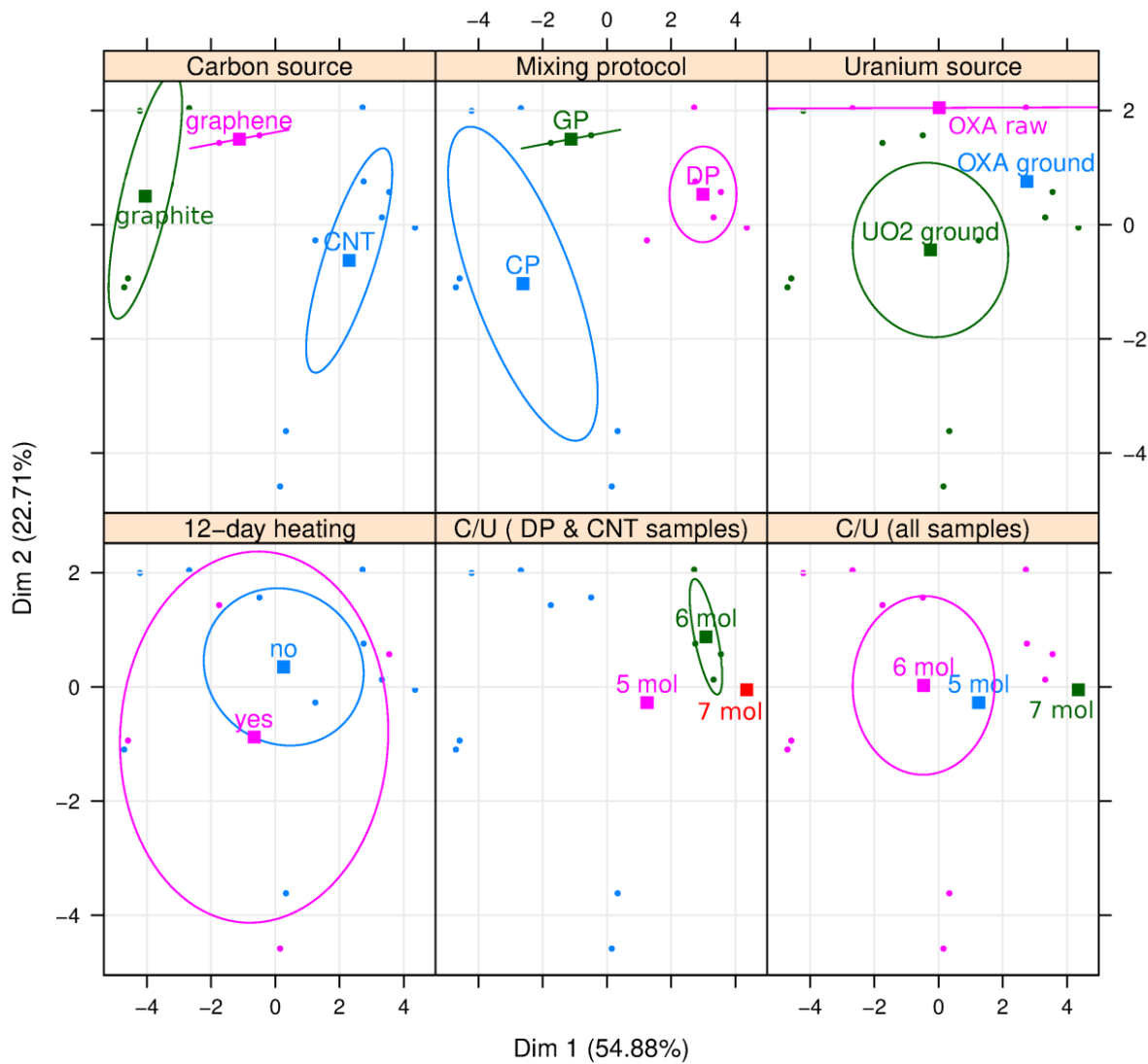

Figure 7: PCA analysis result: qualitative variable graph. The confidence ellipses are drawn around the barycentres (squares) of the samples characterized by one modality of the variable.

Concerning the C/U qualitative variable, when all samples are taken into account, the "5 mol" item lies inside the "6 mol" confidence ellipse. The "6 mol" modality includes very different samples, some elaborated from graphite and graphene, others from CNT and using the CP, DP or GP protocol. The "5 mol" and "7 mol" samples were both prepared with CNT following the DP protocol. In order to highlight the impact of the various modalities of the C/U qualitative variable, it is then more relevant to compare homogeneous groups of samples, in other words to calculate the barycentre and the confidence ellipse from samples made with CNT applying the DP protocol. In this case, the three modalities (5, 6 and 7 mol) differ significantly since they are well separated along Dim 1 and the "5 mol" item is no more inside the "6 mol" confidence ellipse.

In ref. [39], a similar analysis of the qualitative variables was performed: confidence ellipses were drawn around the modalities of each qualitative variables (see Fig. 6 of ref. [39]). The differences between the results obtained in these two analyses serve to highlight the influence of the qualitative variables on the release properties of the samples. For example, the "GP" and "CP" modalities of the "mixing protocol" variable and the "graphene" and "graphite" categories of the "carbon source" variable did not allow us to discriminate between the samples from the physicochemical property analysis. The samples No. 7, 8, 9 and 3, 12 were thus expected to exhibit similar release properties, which was not experimentally observed (see Figs. 4a and 4e). As shown in Figure 7, the "carbon source" and "mixing protocol" variables impact the release properties since the centres of gravity of the "GP" and "CP" modalities on the one hand and of "graphite" and "graphene" on the other hand have different coordinates on the axis strongly correlated to the release properties (Dim 1). Similarly, the difference between the "OXA raw" and "OXA ground" modalities of the "uranium source" variable is much more pronounced when not only the physicochemical properties of the samples but also their release properties are taken into account.

In conclusion, the PCA allowed us to demonstrate that the sample release ability, whatever the considered element, is highly correlated, through the first principal component, to some physicochemical properties of the samples. The samples that best release fission products consist of grains and aggregates with small size and they display a high porosity distributed on small diameter pores. By projecting the samples on Dim 1, a hierarchical ranking is obtained (figure 8) which is based on the most significant variables describing the release ability and the physicochemical properties.

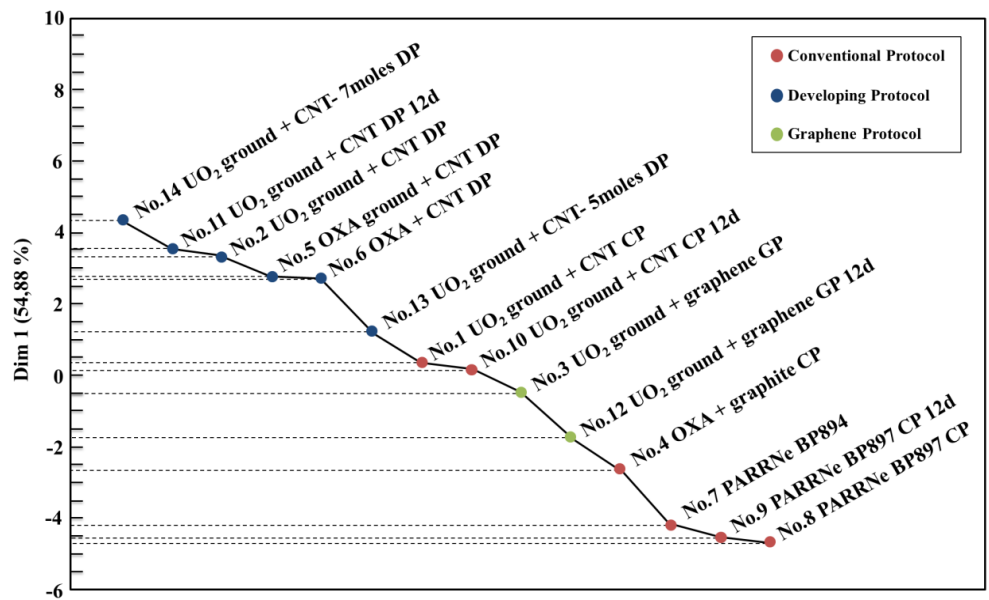

Figure 8: Ranking of the samples according to the PCA "Dim 1".

The sample No. 14 comes first, followed by No. 11, 2, 5 and 6. All these samples, made from carbon nanotubes according to the "DP" protocol, are ranked before ones made with graphite or graphene, and it will be noted that the PARRNe type samples appear ranked among the last ones. The rest of the study will focus on the comparison between the best sample on *RF* (No. 14) and the conventional sample PARRNe (No. 8).

4. Modelling of the expected production rates from an ALTO target

In ISOL facilities, the intensity of radioactive beams produced depends on several parameters, including production in the target and its release efficiency.

In this part, the expected fission number is calculated via the FLUKA code for two targets in the context of the ALTO facility [45], [46]. The two selected targets have the microstructures corresponding to the samples No. 8 and No. 14 and then have different apparent densities. The mechanism for producing the reaction products is the photo-fission induced by the bremsstrahlung radiation emitted by the 50 MeV energy electrons impinging on the target.

The first step consists in modelling the target as well as the oven that contains it. The oven is modelled by a tantalum tube of 20 mm diameter, 200 mm length and 1 mm thickness into which a graphite tube is inserted with 18 mm diameter, 198 mm length and 2.5 mm thickness. The uranium carbide target ($UC_x$ + carbon) is simulated by a cylinder 13 mm in diameter and 193 mm in length placed inside the graphite tube. The target configuration in pellets was not taken into account because it does not affect the production of the fission products but only their release.

In the simulation, the electron beam moves along the z axis and its origin is set arbitrarily 10 mm upstream of the uranium carbide target. Its diameter is 1 cm and its angular distribution is a Gaussian function with 0.3915 cm full width at half maximum, in both x and y directions [47].

Two sets of calculations were performed by simulating the interaction of $2.45\times10^9$ electrons (E = 50 MeV) with uranium carbide targets of 3.82 and 1.27 g.cm$^{-3}$ densities. The first one corresponds to a PARRNe type target (sample No. 8) and the second one to a No. 14 target. Figure 9 shows the electron gamma and fission distributions obtained in each case.

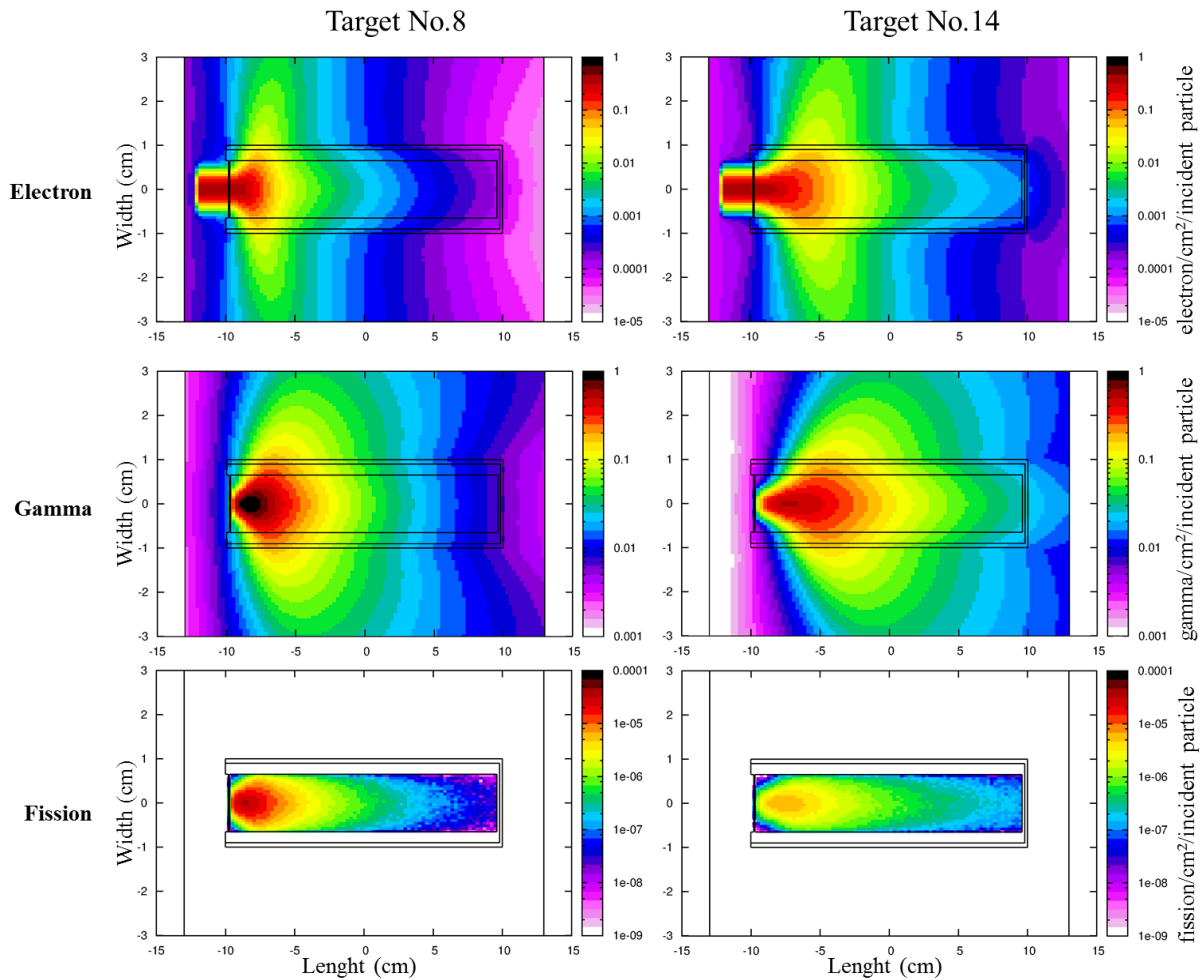

Figure 9: Electron, gamma and fission distributions obtained for No.8 (PARRNe BP897 CP) and No.14 (UO$_2$ ground + CNT – 7 moles DP) targets.

The electrons penetrate more deeply into the target No. 14 than into the target No. 8. About 90% of the electrons interact in the first 5 cm of the target No. 14 but only in the first 3 cm in No. 8. This result was expected because of the target density difference but the calculation quantifies the importance of the phenomenon. The distribution of the fissions also shows a much larger spreading on the z-axis in target No. 14 than in No. 8. But the overall fission number is smaller in the target No. 14 due to its lower density.

In view of these results, a shorter target should be considered at ALTO in order to improve effusion efficiency.

Normalizing the results to a 10 μA electron beam at 50 MeV, i.e. $6.25\times10^{13}$ electrons per second (maximal intensity delivered at ALTO), the number of fissions obtained in the whole target is $2.22\times10^{10}$ for the target No. 8 and

$8.71 \times 10^9$ for No 14. The ratio, 2.55, is lower than the value obtained from the densities of the two targets, which is 3: the best penetration of the beam and the greatest spread of the fissions in the target No. 14 lead to a slight compensation of the deficit expected from the difference in densities. However, in order to compare the performance of both targets, it is important to combine two phenomena: production in the target and release of the fission products.

The diffusion coefficients are extracted from the comparison of the released fractions measured during this work and the values calculated using the formula describing the diffusion process in a cylinder. This formula was obtained by extending the calculations made by Fujioka and Arai in the case of foils, fibers and particles [48]. Assuming isotropic and homogeneous diffusion, the motions along and perpendicular to the cylinder axis are independent then the isotope fraction still remaining in the cylinder, labelled $f(t)$, can be expressed by equation (1) as:

$$f(t) = f(t)_{disk} \times f(t)_{Length} = 4 \sum_{k=1}^{\infty} \frac{exp\left(\frac{-J_{0,k}^2 D}{R^2} t_h\right)}{J_{0,k}^2} \times \frac{8}{\pi^2} \sum_{m=1}^{\infty} \frac{exp\left(\frac{-\pi^2 (2m-1)^2 D}{L^2} t_h\right)}{(2m-1)^2} \quad (1)$$

where $f(t)_{disk}$ and $f(t)_{Length}$ are the isotope fraction still remaining in a disk of radius $R$ (cm) and in an interval of length $L$ (cm) (in other words the radius and thickness of the pellets). $D$ (cm$^2$.s$^{-1}$) is the diffusion coefficient, $t_h$ (s) the heating time and $J_{0,k}$ the kth positive root of the zeroth order Bessel function of the first kind.

From equation (1), $RF$ the fraction released by the cylinder during the heating time $t_h$ is written as equation (2):

$$RF(t_h) = 1 - f(t) = 1 - \frac{32}{\pi^2} \sum_{k,m=1}^{\infty} \frac{exp\left(-Dt_h\left(\frac{J_{0,k}^2}{R^2} + \frac{\pi^2(2m-1)^2}{L^2}\right)\right)}{J_{0,k}^2 (2m-1)^2} \quad (2)$$

The release efficiency ($\varepsilon_{RF}$) of a given isotope of an element [49] can be expressed by equation (3):

$$\varepsilon_{RF}(\lambda) = -\int_0^{\infty} \frac{df(t)}{dt} \times exp(-\lambda t) \, dt = \frac{32}{\pi^2} \sum_{k,m=1}^{\infty} \frac{\pi^2 \frac{(2m-1)^2}{L^2} + \frac{J_{0,k}^2}{R^2}}{J_{0,k}^2 (2m-1)^2 \left(\pi^2 \frac{(2m-1)^2}{L^2} + \frac{J_{0,k}^2}{R^2} + \frac{\lambda}{D}\right)} \quad (3)$$

with $D$, $R$, $L$, $J_{0,k}$ as defined above and $\lambda$ (s$^{-1}$) the radioactive decay constant of the isotope.

Finally, the production rates extracted from the target are related to the number of fissions generated in the target and given by the FLUKA simulation ($N_{fission}$), the fission yield for the given isotope ($Y$) and the release efficiency for this isotope $\varepsilon_{RF}$, according to equation (4):

$$\varphi = N_{fission} \times Y \times \varepsilon_{RF} \quad (4)$$

Figure 10 illustrates the change in the released ratio between the targets No.14 and No.8 as a function of the isotope half-life for the elements measured during the experiment at 1768 °C assuming pellets with 13 mm in diameter and 1.7 mm in thickness.

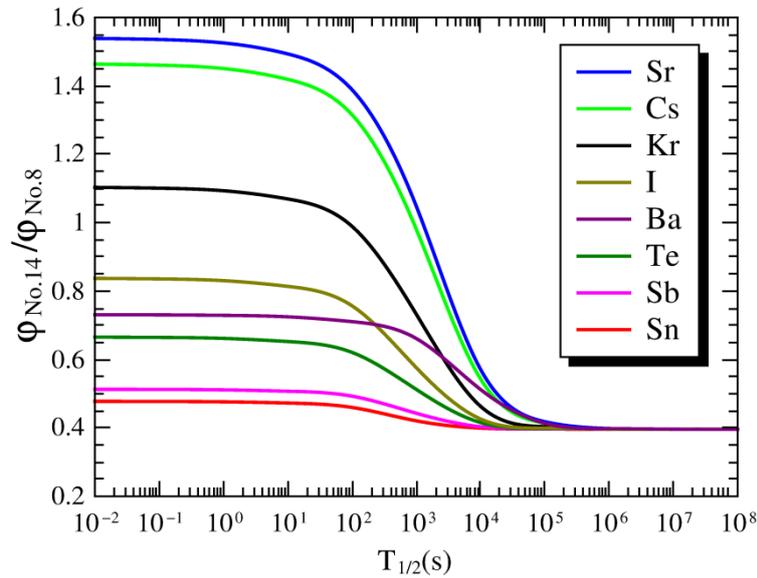

Figure 10: Released ratio between targets No. 14 (UO$_2$ ground + CNT – 7 moles DP) and No. 8 (PARRNe BP897 CP) and as a function of the isotope half-life.

Whatever the element, the ratio $\varphi_{No.14}/\varphi_{No.8}$ is about 0.4 for the long-lived isotopes (T$_{1/2}$ > 10$^5$ s), this figure corresponds to the value of the fission-number ratio between the two targets. It increases for isotopes with shorter half-life, and then remains constant when T$_{1/2}$ < 100 s. The maximum value reached by $\varphi_{No.14}/\varphi_{No.8}$ is around 0.5 for Sn and Sb, elements very well released by both targets, 0.75 for Te, Ba and I, 1.1 for Kr and 1.5 for Cs and Sr. For Kr, the lowest fission number generated in the target No. 14 is counterbalanced by a better release. This compensation phenomenon is not obtained for Ba, Te and I but it is strongly increased for Cs and Sr leading to an amount of short-lived isotopes released 50 % higher by the target No. 14 than No. 8.

5. Conclusion

Some improvements have been implemented in our release-fraction measurement protocol in order to control the reproducibility of the experiments, in particular in the heat treatment step using a programmable logic controller.

To compare the fission-product release efficiency of different samples from different raw precursors, standardization according to the thickness of the pellet is necessary. In these conditions, the released fractions of 11 elements were determined, instead of 13 in our previous works since we demonstrated that the $^{91m}$Y and $^{135m}$Xe should not be considered.

The release from 14 different types of UC$_x$ microstructures was measured. These samples differ by their synthesis parameters: the nature of the carbon precursor (graphite, graphene or CNT), the nature of the uranium precursor (uranium oxide or oxalate) and its grinding, the molar C/U ratio, the mixing protocol (CP, GP, DP) and the long-term heating (12 days, 1800°C). The developed mixing protocol (DP) leads to homogeneous UC$_x$ samples with a porous nanostructure. The nanometric UC$_x$ grains and the micrometric pores, well distributed all together, reduce the diffusion path in the grains and favor the effusion through the porosity, two factors leading to the improvement of the fission-product release.

In order to highlight the correlations between the obtained microstructure and the release efficiency without any experimenter bias; a statistical approach via the principal component analysis (PCA) was performed.

For the PCA, the 14 samples synthesized were described by 17 quantitative and active variables (11 physicochemical variables collected in ref. [39] and 6 release parameters). This analysis demonstrated that these 17 variables can be reduced to only 2, while retaining 77 % of the information contained in the data set. The benefit of a nanostructure made of nanometric UC$_x$ grains and pores is clearly demonstrated. The first principal component is strongly correlated with the release of fission products, total porosity and amount of small-size pores as well as small-size grains. The second principal component is correlated to the Ba and Sr released fractions and the quantity of pores of large size (3 and 30 μm).

The first principal component expressing 55 % of the total dataset inertia, a ranking was done by sorting the samples according to their coordinate along this dimension. The best samples are made with carbon nanotubes following the DP protocol where they were better than those made with graphite or graphene. It can be noted that the PARRNe type samples appear ranked among the last ones. A study on OXA-based samples is still needed to understand their particular behaviour.

A modelling of the on-line ALTO conditions was performed using the FLUKA code on the best target (No. 14) and on a PARRNe one for comparison. The results showed that the less dense target (No. 14) allows deeper penetration of the beam and therefore a larger distribution of fissions. Nevertheless, this new target has a production rate three times lower than a conventional one. However, a better release for all elements was measured experimentally, leading to a significant production gain for short-lived isotopes ($T < 100$ s) of Cs and Sr.

This study shows that there is no universal target material microstructure for the release of isotopes and that the microstructure of the target must be adapted according to the elements to be released.

Acknowledgements

The authors would like to thank Ulli Köster, Pascal Jardin and Olivier Tougait for their comments and fruitful discussions. We would like to thank the ALTO team for the help in the implementation of the experimental devices. We would like to thank the NESTER, SPR and ALTO teams for advice and discussion on this project and especially Abdelhakim Saïd, Thony Corbin, Christophe Planat, Robert Leplat, Christophe Vogel, Alain Semsoum, Sébastien Wurth and David Verney for their help during the experiments.

**Complementary data** [1]**:**

Table 1: Summary of the samples with the different qualitative variables used

| Samples | 12-day heating | C/U | Carbon source | Uranium source | Mixing protocol |
|---|---|---|---|---|---|
| No.1 $UO_2$ ground + CNT CP | no | 6 | CNT | $UO_2$ ground | Conventional Protocol |
| No.2 $UO_2$ ground + CNT DP | no | 6 | CNT | $UO_2$ ground | Developing Protocol |
| No.3 $UO_2$ ground + graphene GP | no | 6 | graphene | $UO_2$ ground | Graphene Protocol |
| No.4 OXA + graphite CP | no | 6 | graphite | OXA | Conventional Protocol |
| No.5 OXA ground + CNT DP | no | 6 | CNT | OXA ground | Developing Protocol |
| No.6 OXA + CNT DP | no | 6 | CNT | OXA | Developing Protocol |
| No.7 PARRNe BP894 | no | 6 | graphite | $UO_2$ ground | Conventional Protocol |
| No.8 PARRNe BP897 CP | no | 6 | graphite | $UO_2$ ground | Conventional Protocol |
| No.9 PARRNe BP897 CP 12d | yes | 6 | graphite | $UO_2$ ground | Conventional Protocol |
| No.10 $UO_2$ ground + CNT CP 12d | yes | 6 | CNT | $UO_2$ ground | Conventional Protocol |
| No.11 $UO_2$ ground + CNT DP 12d | yes | 6 | CNT | $UO_2$ ground | Developing Protocol |
| No.12 $UO_2$ ground + graphene GP 12d | yes | 6 | graphene | $UO_2$ ground | Graphene Protocol |
| No.13 $UO_2$ ground + CNT-5moles DP | no | 5 | CNT | $UO_2$ ground | Developing Protocol |
| No.14 $UO_2$ ground + CNT-7moles DP | no | 7 | CNT | $UO_2$ ground | Developing Protocol |

Table 2: Summary of the carburized samples with the different quantitative variables used. $P_{35}$, $P_{200}$, $P_{3000}$, $P_{10000}$ and $P_{30000}$ represent the percentages of open porosity on pores with diameters 0.035 µm, 0.2 µm, 3 µm, 10 µm and 30 µm, respectively.

| | XRD* | | | | | BET | SEM | He Pycnometry | | Hg Porosimetry | | | | |
|---|---|---|---|---|---|---|---|---|---|---|---|---|---|---|
| | Phase and proportion (%, ± 1 %) | | | Crystallite size (nm, ± 5 nm) | | $UC_x$ Grain size (nm)** | $UC_x$ Aggregate size (µm) | Porosity (%, ± 1 %) | | Open pore size Distribution (%) | | | | |
| | UC | $UC_2$ | C | UC | $UC_2$ | | | Open | Close | $P_{35}$ | $P_{200}$ | $P_{3000}$ | $P_{10000}$ | $P_{30000}$ |
| No.1 $UO_2$ ground + CNT CP | 3 | 88 | 9 | 59 | 87 | 118 | 15 | 78 | 7 | 22 | 10 | 10 | 0 | 58 |
| No.2 $UO_2$ ground + CNT DP | 5 | 86 | 9 | 39 | 114 | 100 | 0.5 | 68 | 12 | 34 | 32 | 34 | 0 | 0 |
| No.3 $UO_2$ ground + graphene GP | 4 | 88 | 8 | 51 | 129 | 1200 | 18 | 49 | 7 | 3 | 22 | 56 | 0 | 19 |
| No.4 OXA + graphite CP | 4 | 87 | 9 | 55 | 160 | 820 | 23 | 55 | 5 | 2 | 12 | 86 | 0 | 0 |
| No.5 OXA ground + CNT DP | 13 | 78 | 9 | 40 | 127 | 94 | 0.6 | 70 | 15 | 24 | 29 | 47 | 0 | 0 |
| No.6 OXA + CNT DP | 7 | 84 | 9 | 65 | 149 | 82 | 3.2 | 74 | 14 | 17 | 17 | 66 | 0 | 0 |
| No.7 PARRNe BP894 | 5 | 86 | 9 | 102 | 165 | 906 | 31 | 41 | 5 | 4 | 14 | 82 | 0 | 0 |
| No.8 PARRNe BP897 CP | 5 | 87 | 8 | 38 | 145 | 972 | 65 | 51 | 5 | 2 | 8 | 46 | 44 | 0 |
| No.9 PARRNe BP897 CP 12d | 5 | 87 | 8 | 46 | 144 | 914 | 56 | 49 | 8 | 2 | 5 | 49 | 44 | 0 |
| No.10 $UO_2$ ground + CNT CP 12d | 3 | 88 | 9 | 48 | 86 | 100 | 71 | 72 | 13 | 19 | 13 | 10 | 0 | 58 |
| No.11 $UO_2$ ground + CNT DP 12d | 5 | 86 | 9 | 42 | 110 | 96 | 0.5 | 64 | 17 | 30 | 33 | 37 | 0 | 0 |
| No.12 $UO_2$ ground + graphene GP 12d | 4 | 88 | 8 | 43 | 135 | 1412 | 14 | 48 | 4 | 3 | 17 | 65 | 0 | 15 |
| No.13 $UO_2$ ground + CNT-5moles DP | 5 | 90 | 5 | 57 | 119 | 104 | 0.6 | 64 | 8 | 29 | 27 | 35 | 0 | 9 |
| No.14 $UO_2$ ground + CNT-7moles DP | 4 | 84 | 12 | 40 | 102 | 92 | 0.2 | 69 | 15 | 33 | 37 | 30 | 0 | 0 |
| Standard deviation | 2 | 3 | 1 | 16 | 24 | 484 | 26 | 11 | 4 | 12 | 10 | 22 | 15 | 20 |

* For all the samples, the agreement factors of the Rietveld refinement were in the ranges: 11.6% < $R_w$ < 14.8%, 5.8% < $R_{exp}$ < 7.3%, 1.9 < $\chi^2$ < 2.3.
** Error bar of SSA measurements is 5%

The crystallite sizes, obtained by XRD after Rietveld analysis, are approximately the same for all samples.

For the SSA analysis, the UC$_x$ grains cannot be considered as isolated (due to sintering leading to aggregation even agglomeration of grains) contrary to what is assumed for a standard SSA analysis. Moreover, a roughness of the grain surface increases its specific surface. These parameters could overestimate the average size deduced [2]. The UC$_x$ grains were assumed spherical, the average size was obtained taking into account the excess-carbon contribution as explained in Part 1.

Aggregate size estimation is more delicate and has been carried out using the intercept technique [3].